\documentclass{aastex}

\usepackage{emulateapj5}
\usepackage{latexsym}

\newcommand \kms          {\hbox{km~s$^{-1}$}}
\newcommand \msol         {\hbox{M$_{\odot}$}}           % Solar mass

\newcommand \ha           {H$\alpha$}

\newcommand \nii          {[\ion{N}{2}]}
\newcommand \vhel         {\hbox{V$_{\rm hel}$}}           % Solar mass

\begin{document}
\submitted{Accepted for publication in the Astronomical Journal}
\title{Variable \ha\ Line Emission from the Central Star of the Helix Nebula}
\author{Robert A.\ Gruendl, You-Hua Chu\altaffilmark{1}, Ian J.\ O'Dwyer, and 
Mart\'{\i}n A.\ Guerrero}
\affil{Astronomy Department, University of Illinois, 
1002 West Green Street, Urbana, IL 61801}
\email{gruendl@astro.uiuc.edu, chu@astro.uiuc.edu, iodwyer@astro.uiuc.edu,
mar@astro.uiuc.edu}
\altaffiltext{1}{Visting astronomer, Cerro Tololo Inter-American Observatory}

%\received{}
%\accepted{}
%\journalid{}{}
%\articleid{}{}

\begin{abstract}

The central star of the Helix Nebula is a hot white dwarf with 
unexpected hard X-ray emission.  With an effective temperature
of $\sim$100,000 K, the star is a powerful source of H-ionizing
radiation; the atmosphere of a stellar or planetary companion,
if present, will be ionized and emit recombination lines.  
To probe the origin of hard X-ray emission from the Helix 
Nebula's central star, we have obtained multi-epoch 
high-dispersion spectra of the star, and found temporal 
variation in the H$\alpha$ line profile over a time span
of one week.  The observed width and strength of the variable
\ha\ emission component are consistent with the hypothesized dMe 
companion proposed to explain the hard X-ray emission.
A dMe companion, however, cannot explain the possible detection of 
variable \ion{He}{2} and \nii\ emission. Follow-up spectroscopic 
monitoring of the Helix Nebula central star is needed to better 
establish the identification of the spectral lines and their 
temporal behavior in order to determine the origin of the
optical variability and hard X-ray emission.

\end{abstract}


\section{Introduction}

The central star of the Helix Nebula (NGC\,7293) is a hot white dwarf, 
WD\,2226$-$210, with an effective temperature of
$\sim$100,000 K \citep{mendez88,napi99}.  X-ray observations of 
the Helix Nebula using {\it ROSAT} and {\it Chandra} show a point source 
at the position of the central star \citep{leahy94,guerrero01}.
This X-ray source not only shows a soft spectral component from the 
photosphere, but also a harder spectral component whose origin is 
uncertain.  A hot white dwarf by itself is not expected to emit hard
X-rays; the hard X-ray spectral component must originate externally.
It has been suggested that this hard X-ray emission
is caused by the interaction between a fast stellar wind and the
ambient nebular material \citep{leahy96}, although the central star
of the Helix Nebula is not known to have a fast stellar wind \citep{pp91} and
the X-ray emission is not diffuse as seen in the Cat's Eye Nebula
\citep{chu2001b}.  The Helix Nebula central star is not known to have a 
binary companion \citep{ciar99} and has been used as a standard 
star \citep{Oke90}.  What is the origin of the hard X-ray
emission from the central star of the Helix Nebula?

An excellent opportunity to search for clues to the origin of this
hard X-ray emission is provided by the Helix Nebula central star itself.
As a 100,000 K hot white dwarf, it is a powerful source of H-ionizing
UV radiation.  Any gaseous objects in its close vicinity will be ionized 
and produce H$\alpha$ line emission.  
High-dispersion (R $\ge$ 30,000), long-slit observations of this 
H$\alpha$ emission can be used to determine its spatial extent, 
velocity profile, and temporal variations, which supply pertinent 
information about the immediate surroundings of the star.
Spatially extended H$\alpha$ emission indicates a nebular origin.
Spatially unresolved H$\alpha$ emission could originate from the 
atmosphere of the white dwarf itself or the ionized
atmosphere of a companion star or a Jovian planet.  These possibilities
could be distinguished from the orbital velocity and period measurements
of the H$\alpha$ line \citep{chu2001a}.  High-dispersion spectroscopy 
of hot white dwarfs has been used to demonstrate the existence of dMe
companion stars \citep{vennes99,kawka2000}.

To probe the environment of the central star of the Helix Nebula, we have 
obtained multi-epoch, long-slit, high-dispersion spectra at the H$\alpha$ 
and \nii\ $\lambda\lambda$6548,6583 lines.  Temporal variations
of the spectra are detected!  In this paper we report the observations
and discuss the significance of our results.

\section{Observations and Reduction}

We obtained high-dispersion spectroscopic observations of the 
central star of the Helix Nebula with the echelle spectrograph 
on the 4m telescope at the Cerro Tololo Inter-American Observatory 
on 2000 December 4, 8, and 11.  These observations were made during
two unrelated programs, and thus used two different instrumental 
configurations (chosen for those programs).

In each observation a 79 line mm$^{-1}$ echelle grating was used.  The 
observations on 2000 December 4 were made in a multi-order mode, 
using a 226 line mm$^{-1}$ cross-disperser
and a broad-band blocking filter (GG385), while the observations on
2000 December 8 and 11 were made in a single-order mode, using a 
flat mirror and a broad H$\alpha$ filter (central wavelength 6563 \AA\
with 75 \AA\ FWHM) to isolate the order containing H$\alpha$.  
For each observation the long-focus red camera was used to obtain a 
reciprocal dispersion of 3.5 \AA\ mm$^{-1}$ at H$\alpha$.  
The spectra were imaged using the SITe2K \#6 CCD detector.  
The 24 $\mu$m pixel size corresponds to 0\farcs 26 pixel$^{-1}$ along 
the slit and $\sim$0.08 \AA\ pixel$^{-1}$ along the dispersion axis.  
The observations on December 4 used a 1\farcs 0 slit, while the 
observations on December 8 and 11 used a 1\farcs 6 slit.  The 
resultant instrumental resolution, as measured by the FWHM of the
unresolved telluric emission lines, was 0.24 \AA\ and 0.29 \AA\
(11 \kms\ and 13 \kms\ at H$\alpha$) for these two slit widths, 
respectively.

The observations of the Helix Nebula central star are summarized
in Table~\ref{obs_sum}.  Two exposures were obtained at each epoch to
aid in cosmic-ray removal.  The observations were reduced using 
standard packages in $IRAF$\footnote{IRAF is distributed by the National
Optical Astronomy Observatories, which are operated by the Association
of Universities for Research in Astronomy, Inc., under cooperative 
agreement with the National Science Foundation}.
All spectra were bias and dark corrected, and cosmic-ray events were 
removed by hand.  Observations of a quartz lamp were used to correct
for pixel-to-pixel sensitivity variations in the CCD.  Observations 
of a Th--Ar lamp were used to determine the dispersion correction
and then telluric lines in the source observations were used to check
for the absolute wavelength calibration.

\section{Results and Analysis}

\begin{figure}
\epsscale{0.4}
\plotone{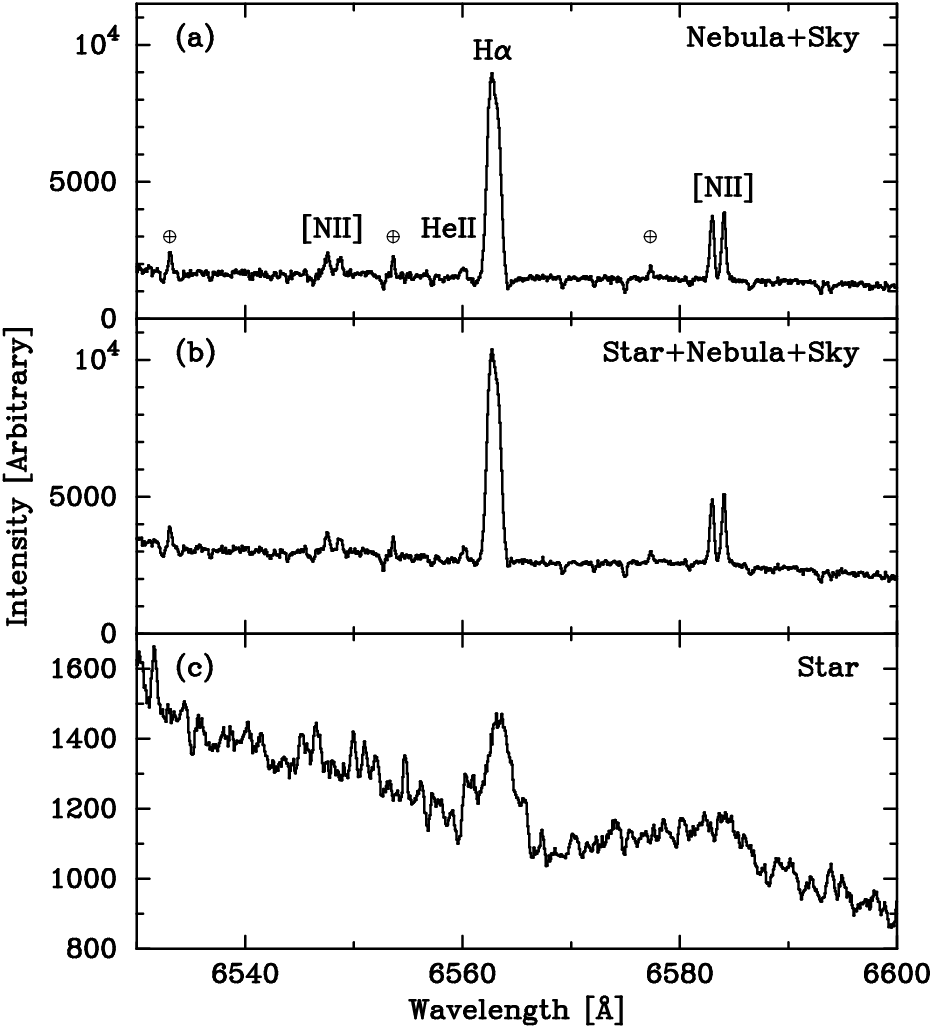}
\caption{Long-slit echelle observations of the Helix Nebula central 
star from 2000 December 11.  (a) Nebular spectrum extracted from 
positions adjacent to the Helix Nebula central star, (b) stellar spectrum 
extracted with an aperture centered on the central star, and 
(c) stellar spectrum after the nebular and sky background has been
subtracted.  Note that the only spectral response that has been 
removed is the falloff due to the echelle blaze.  The telluric OH lines
are identified with a ``$\oplus$''.}
\label{neb_subtract}
\end{figure}

The spectrum of the central star is contaminated by nebular and
sky emission.  In order to investigate the detailed spectral 
properties of the central star this contamination must be subtracted.
This can be accomplished because the nebular and sky emission components
are both spatially extended while the stellar emission is unresolved.  
In Figure~\ref{neb_subtract} we show the raw spectra extracted from the 
echellograms acquired on 2000 December 11.  The observations at this
epoch are the most heavily contaminated by a 
solar component due to the nearly full lunar phase.
Figures~\ref{neb_subtract}a\&b present the raw spectra 
obtained off and on the star, respectively.  These spectra show a number
of telluric OH lines, solar absorption features (most prominently
between 6565 \AA\ and 6575 \AA ), and the H$\alpha$ and \nii\
$\lambda\lambda$6548,6583 nebular emission lines from the planetary nebula.
The difference between these two spectra, the ``background-subtracted'' 
stellar spectrum, is shown in Figure~\ref{neb_subtract}c.

The two different instrument configurations, multi- and single-order, 
result in different spectral responses.
These spectral responses must be accurately determined
in order to compare the observations made at different epochs.
Observations of the spectrophotometric standard star HR\,3454 were 
obtained for flux calibration, but the available spectrophotometry
for this standard star \citep{hamuy92} has too coarse a
resolution, 16 \AA, to correct for the 
complex passband shape which arises from the combination of the 
echelle blaze and narrow-band H$\alpha$ filter used for the 
single-order echelle observations.  
To circumvent these spectral response problems we have normalized
our observations to the stellar continuum using a polynomial fit 
guided by the 2000 December 4 observations that were obtained with 
the multi-order echelle mode.  
These continuum fits constitute the largest source of uncertainty 
when comparing the background-subtracted stellar spectra.

\begin{figure}
\epsscale{0.4}
\plotone{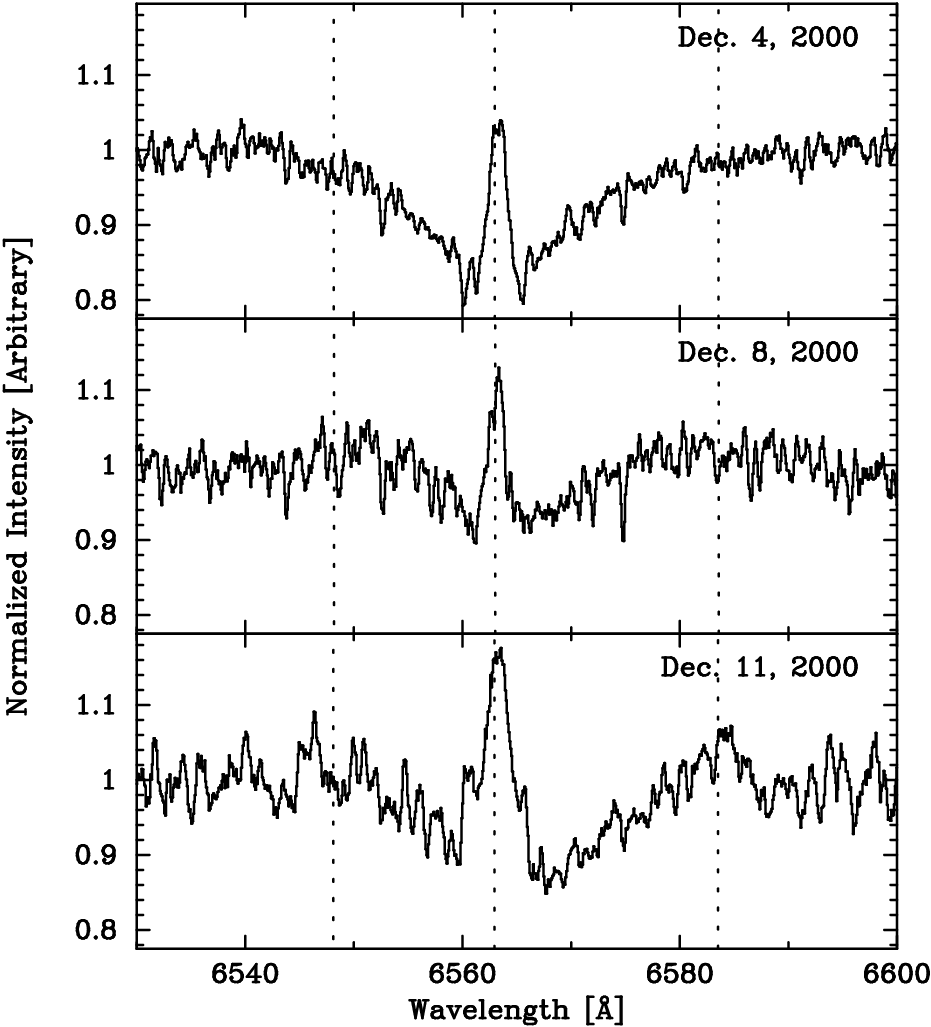}
\caption{The continuum--normalized stellar spectrum near H$\alpha$ 
for the hot white dwarf central star of the Helix Nebula at three epochs.
A clear change in the H$\alpha$ and possibly in the \ion{He}{2} $\lambda$6560
and [\ion{N}{2}] $\lambda$6583 emission-line profiles are detected in 
the 2000 December 11 observations.
The nebular systemic velocity, as measured from the [\ion{N}{2}] 
$\lambda$6583 line, is indicated by dashed vertical lines for H$\alpha$
and [\ion{N}{2}] $\lambda\lambda$6548,6583.
\label{helix_spec}
}
\end{figure}

The continuum--normalized, background--subtracted spectra of the 
Helix Nebula central star have a S/N of $\sim$35 in each 0.08 \AA\ 
pixel.  To improve the S/N ratio, we have smoothed the spectra with a 
5-pixel boxcar filter.  The smoothed spectra are presented in
Figure~\ref{helix_spec}.  Note that the S/N in the smoothed
spectra does not appear to be the expected $\sim$80 per 0.4 \AA\ 
resolution element because the spectra contain systematic effects 
such as telluric absorption lines and uncertainties in the
spectral responses.

The stellar line profile is complex and may be variable; therefore,
we have used the nebular lines (see Figures~\ref{neb_subtract}a\&b)
to define a reference frame, and will use heliocentric velocity (\vhel)
throughout the remainder of this paper.  
Because the nebular H$\alpha$ line is blended with the telluric 
\ha\ emission we have used the \nii\ $\lambda$6583 line for this analysis, 
adopting a rest wavelength of $\lambda$6583.454 \AA\ \citep{Spyr95}.   
We find that the nebular emission has two components,
at $0\pm$2 and $-$49$\pm$2 \kms, 
and therefore we assume the systemic velocity of the nebula is the 
average of these two velocities, roughly $-$24 \kms.  
Since the nebular material originates from the
white dwarf, we may assume that the time-averaged \vhel\ for the 
white dwarf (over a long period of time) is also roughly $-$24 \kms.

Over the week-long baseline for these observations the H$\alpha$ 
and possibly \ion{He}{2} $\lambda$6560 and \nii\ $\lambda$6583 lines 
showed variation. 
In all stellar spectra there was a ``narrow'' H$\alpha$ emission component
with \vhel$=-$10$\pm$5 \kms\ and FWHM of 80 \kms\ superposed on the
broad H$\alpha$ absorption from the white dwarf atmosphere.  The broad
absorption is centered at \vhel$=-$6$\pm$10 \kms\ slightly redder than the
``narrow'' emission component.  In the 2000 December 11 spectrum 
the H$\alpha$ line profile changed dramatically and showed an additional 
broad emission component superposed on the ``narrow'' emission component. 
 
\begin{figure}
\epsscale{0.4}
\plotone{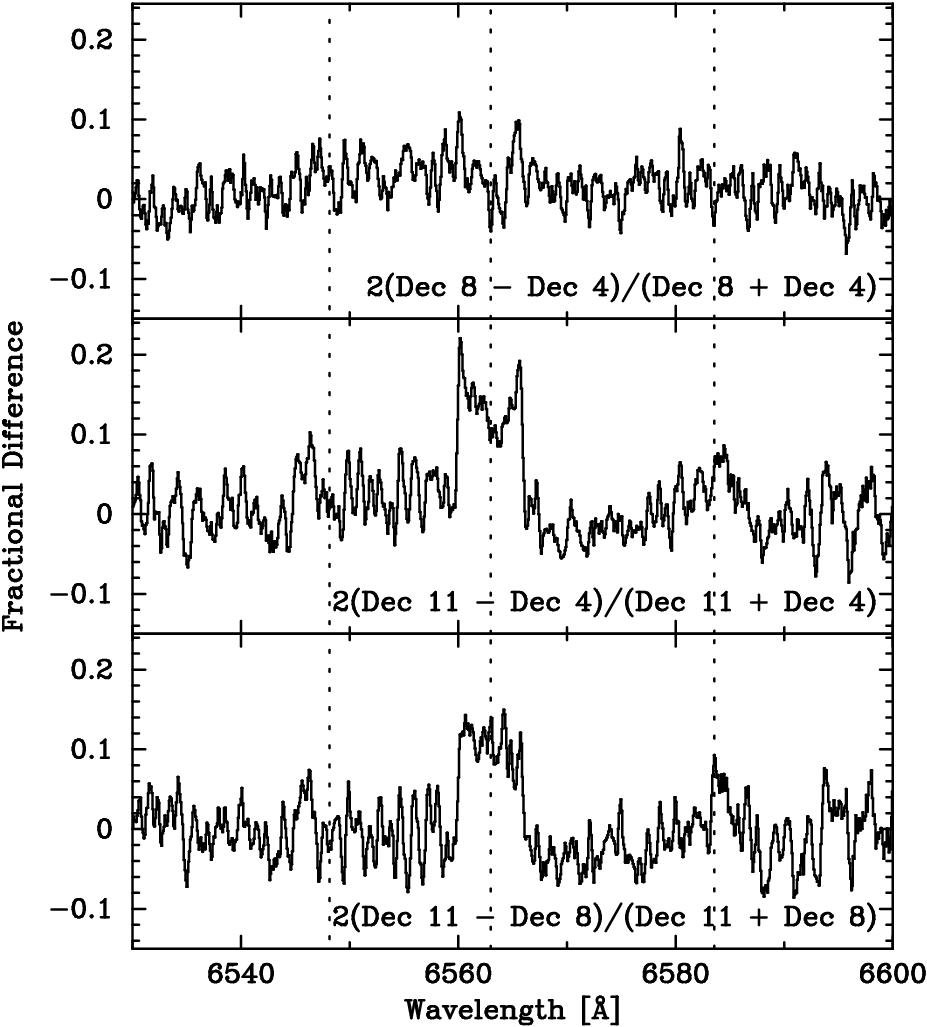}
\caption{Fractional difference of the continuum-normalized spectra
for the three observed epochs.  
The nebular systemic velocity is indicated by dashed vertical lines
as in Figure~\ref{helix_spec}.
\label{helix_lincomb}
}
\end{figure}

\begin{figure}
\epsscale{0.4}
\plotone{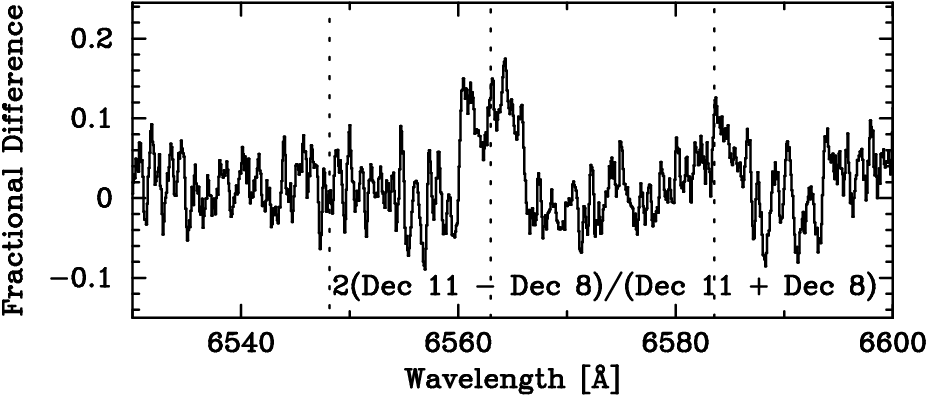}
\caption{Fractional difference of the unnormalized spectra from
2000 December 8 and 11 (the two observations with the same observational
setup).  Little qualitative difference is seen when compared to 
bottom panel in Figure~\ref{helix_lincomb}.  As in 
Figures~\ref{helix_spec}\&\ref{helix_lincomb} the
nebular systemic velocity is indicated by dashed vertical lines.
\label{helix_nonorm}
}
\end{figure}

To better highlight the quantitative difference from one epoch to 
another we have calculated the fractional difference spectrum, 
$2\times({\rm epoch}_m(\lambda)-{\rm epoch}_n(\lambda))/
({\rm epoch}_m(\lambda)+{\rm epoch}_n(\lambda))$, for each epoch pair. 
In these fractional difference spectra, shown in Figure~\ref{helix_lincomb},
the narrow emission component does not appear to vary (i.e., is not
present) while the variable emission component has a broad, 
flat-topped spectral shape with roughly 10\% the brightness of 
the continuum.  Careful examination of the 
spectra also show that there are differences between the wings of 
the broad H$\alpha$ absorption line.  In order to determine whether
these features are real or the result of the artifacts from the
continuum normalization, we have calculated the fractional difference
spectrum without normalizing the continuum for the epochs with 
the same instrumental setup, i.e., December 8 and 11.  This fractional 
difference spectrum is presented in Figure~\ref{helix_nonorm} and shows
that the low-level ($<$ 4\%) variations in the broad wings of the white
dwarf H$\alpha$ absorption are likely real, rather than an artifact from 
the continuum normalization.
This conclusion is further strengthened by a similar analysis carried 
out for the observations of the standard star HR\,3454 made in these 
two epochs, which show no variation in the fractional difference spectrum 
within a part in 10$^4$.

If the broad, flat-topped variable component is \ha\ emission, it is 
centered at \vhel$=-$18$\pm$10 \kms\ and has a full width of $\sim$315 \kms.  
As the \ion{He}{2}~$\lambda$6560 line is close to the \ha\ line, 
this broad emission component may also contain \ion{He}{2} emission.
The possibility that there is a \ion{He}{2} component in the stellar 
spectrum is best demonstrated in Figure~\ref{neb_subtract}, where the 
blue shoulder of the stellar \ha\ emission (Figure~\ref{neb_subtract}c)
can be compared to the nebular \ion{He}{2} emission 
(Figures~\ref{neb_subtract}a\&b).
%If the blue shoulder of the broad \ha\ emission is \ion{He}{2} emission, 
%then it is red-shifted from the nebular \ion{He}{2} line by at least 24 \kms.  
The centroids and widths of this possible \ion{He}{2} component and 
remaining \ha\ component may be measured from the fractional difference
spectrum in Figure 4.

Finally, we note that a feature consistent with \nii\ $\lambda$6583 
emission but centered at \vhel$=+$15$\pm$10 \kms\ with a FWHM of 
105$\pm$20 \kms\ appeared in the 2000 December 11 spectrum (see Figure 4).
The weaker \nii\ $\lambda$6548 line, however, was not detected.
The velocities of all nebular and stellar lines are summarized in 
Table~\ref{kin_sum}.   

\section{Discussion}

Previous $ROSAT$ observations of the Helix Nebula show an unresolved
X-ray source coincident with the central star.  The spectrum of this
source has a soft component ($<$0.2~keV) consistent with emission 
from the 100,000~K photosphere of the white dwarf and a hard component
which peaks at $\sim$0.8 keV \citep{leahy94}.  Recent higher spatial
resolution $Chandra$ observations by \citet{guerrero01} confirm that
the emission in this hard X-ray component originates from an unresolved 
source coincident with the central white dwarf to within 0\farcs 5.
Furthermore, the X-ray 
flux of this point source showed variations over a $\sim$70 ksec time 
span with 3$\sigma$ significance.  They discuss the possible origins 
of the X-ray emission and conclude that the X-ray emission is likely 
due to coronal activity from an unseen dMe (dwarf M star with emission
lines) companion.  

Our high-dispersion echelle spectra of the central star of the Helix 
Nebula show constant broad H$\alpha$ absorption superposed with a 
narrow emission component.  Such an \ha\ profile has been commonly 
seen in hot white dwarfs \citep{reid89} and the emission component 
arises due to non-LTE effects in the white dwarf atmosphere 
\citep{lanz95,hubeny99}.

Our observations also show, for the first time, variability in the 
optical spectrum of the central star of the Helix Nebula.  
This variable spectral component, spatially unresolved and
coincident with the stellar spectrum, includes clear variability
in the H$\alpha$ emission and 
possible variability in the \ion{He}{2} and \nii\ $\lambda$6583 emission.  
Due to the close proximity of the H$\alpha$ and \ion{He}{2} $\lambda$6560
lines and the broad width of the variable emission component, we can not
rule out the possibility that the variable emission is a combination of
H$\alpha$ and \ion{He}{2} emission.  

Are our observations consistent with H$\alpha$ emission from the
atmosphere of a dMe stellar companion?  
Using an H$\alpha$ flux scaled from the H$\gamma$ emission 
line strength of the prototypical dMe star YZ CMi during a flare state
\citep{doyle88} and the flux density of the central star of the Helix 
Nebula \citep{Oke90}, we find the observed intensity of the variable 
\ha\ emission is consistent with that of a flaring dMe star.
It must be noted, however, that the observed flux in the variable 
H$\alpha$ emission alone cannot distinguish among (1) Jovian planets with 
fluorescing atmospheres at small distances from the hot white dwarf, 
(2) dwarf companion stars with fluorescing atmospheres at larger 
distances from the hot white dwarf, and (3) \ha\ emission which 
arises from a dMe companion star of the white dwarf.  

An additional constraint is provided by the large width of the variable 
emission line.  If this line is entirely due to \ha\ emission from a
flourescing planetary or stellar companion, the observed width (315 \kms)
is too large to be explained  by orbital or 
rotational motion.  On the other hand, dMe stars during flares have been
observed to show emission line variability with line widths as large as 
300 \kms\ \citep{doyle88,eason92}.  Furthermore, the X-ray flux
of such stars in their quiescent state \citep{white96} is similar to
the flux observed from the Helix Nebula central star \citep{guerrero01}.  
Observations by \citet{ciar99} constrain that such a companion must 
have a spectral type later than M5.  It seems that a late-type 
dMe companion is the best candidate to explain all observations
of the Helix Nebula central star, with the X-ray observations during a 
quiescent state and the broad \ha\ emission from a flare.  
A dMe companion, however, would not be able to explain the variable 
\ion{He}{2} and \nii\ emission, if these lines are identified 
correctly and truly detected.  Since the presence of these lines
is by no means certain, a dMe companion remains the most viable 
candidate to explain all well established observations of the Helix
Nebula central star.

We may further examine whether the observed velocity and 
variation of the broad H$\alpha$ component of the proposed
orbiting dMe companion are consistent with the orbital
distance constrained by the optical and X-ray sizes.
The size and position of the X-ray source measured with 
the $Chandra$ ACIS-S image requires that the dMe companion
must orbit within 0\farcs5 from the Helix Nebula central star.
At a distance of 210 pc \citep{harris97}, 0\farcs5 corresponds to
105 AU.  A 0.2 M$_\odot$ dMe star orbiting at 105 AU from
a 0.6 M$_\odot$ PN central star would have an orbital 
velocity of 2.6 km~s$^{-1}$ and an orbital period of 1,200 yr.
The observed velocity of the broad H$\alpha$ emission does
not conflict with this orbital velocity.

If the broad variable emission line is assumed to be a composite 
of roughly equal strength redshifted \ha\ and \ion{He}{2} emission 
then the measurements on 11 December imply that a flourescing 
planetary or stellar companion may be viable candidates to explain 
the variable \ha\ emission.  In this case the centroid of the portion
of the variable emission that would be assigned to \ha\ implies that
the orbital velocity of a flourescing planetary or stellar companion 
is at least 80 \kms.
For a Jupiter mass companion orbiting a 0.6 \msol\ white dwarf this
places upper limits on the semi-major axis and period of $\sim$0.08 AU 
and $\sim$11 days.  For a 0.2 \msol\ stellar companion the upper limits
on the semi-major axis and period would be $\sim$0.11 AU and $\sim$15 days. 

Clearly, our detection of variability of the emission lines in the 
spectrum of the central star of the Helix Nebula has provided useful
constraints on the possible origins of its X-ray emission.
The limited number and moderate S/N of the observations presented here 
do not allow us to prove beyond a reasonable doubt that a dMe companion 
is a unique solution.  
Follow-up spectroscopic monitoring of the central star in the Helix 
Nebula with high-dispersion (R$\ge$30,000), and high S/N is needed to 
better identify the variable spectral features, and to establish their 
temporal behavior.  
Measurement of the temporal behavior at just a few more epochs will 
quickly discriminate between flourescent emission (from either a planetary
or a stellar companion) and emission that arises from a dMe companion.
Other observations that might prove useful include, near-IR photometry
to determine whether or not an IR excess from a low--mass stellar companion 
is present.  

It is important to confirm the binarity of the Helix Nebula central star.
The Helix Nebula is one of the closest PN yet all prior attempts to search 
for a binary companion (spectroscopic, photometric and $HST$ WFPC2 imaging)
have found no evidence for a binary companion.  It has been suggested that 
binary companions play significant roles in the aspherical mass loss of PN
central stars, and thus are responsible for the wide range of PN
morphologies observed.
Our results suggest that high-dispersion spectroscopic observations
may provide a more sensitive probe for stellar companions in PNs.
A large sample of PNe with a wide range of morphologies need to be 
observed so that we may assess the importance of binary central stars
on the formation and evolution of PNs.

\acknowledgements
We would like to thank C. Pilachowski, H. Bond, E. Sion, and J. Holberg 
for helpful discussions, and R. Williams and C. Danforth for agreeing 
to contribute telescope time to this project.
R.A.G. is supported by the NASA grant SAO GO 0-1004X.  
M.A.G. acknowledges the support by the Direcci\'on General de
Ense\~nanza Superior e Investigaci\'on Cient\'{\i}fica of 
Spanish Ministerio de Educaci\'on y Cultura.

\clearpage

\clearpage

\begin{deluxetable}{lcl}
\tablewidth{0pt}
\tablecaption{Observation Summary\label{obs_sum}}
\tablehead{
\colhead{Median Date and Time} & \colhead{Exposure Times} &
\colhead{Echelle Setup}\\
\colhead{[UT]} & \colhead{[seconds]} & 
\colhead{}}
\startdata
Dec. 04, 2000 -- 01:17:56 & 2$\times$900  & multi-order \\
Dec. 08, 2000 -- 02:05:08 & 2$\times$1200 & single-order \\
Dec. 11, 2000 -- 00:43:03 & 2$\times$1200 & single-order \\
\enddata
\end{deluxetable}

\begin{deluxetable}{llcr}
\tablewidth{0pt}
\tablecaption{Summary of Kinematic Components\label{kin_sum}}
\tablehead{
\colhead{}          & \colhead{}       & \colhead{FWHM}         & \colhead{V$_{\rm Hel}$} \\
\colhead{Component} & \colhead{Remark} & \colhead{[km~s$^{-1}$]} & \colhead{[km~s$^{-1}$]}
}
\startdata
\nii\ $\lambda$6583 & nebular, approaching &       20  &  0~$\pm$2\phn        \\
                    & nebular, receding    &       20  &  $-$49~$\pm$2\phn    \\
                    & nebular, systemic    &  \nodata  &  $-$24~$\pm$2\phn  \\
                    &              &           &                      \\
\ha\ (absorption)\tablenotemark{a} &  constant &  \nodata  &  $-$6~$\pm$10      \\
\ha\ (emission)     &  constant    &  \phn 80  &  $-$10~$\pm$5\phn  \\
                    &              &           &                      \\
\hline
                    &              &           &                      \\
\ha\ (emission)\tablenotemark{b} & variable &      315  &  $-$18~$\pm$10     \\
~~~~~~~~~~or           &              &           &                      \\
\ha\ (emission)\tablenotemark{c}            & variable &      155  &  $+$24~$\pm$10     \\
\ion{He}{2} (emission)\tablenotemark{c}  & variable & \phn 60 &  $+$17~$\pm$10    \\
                    &              &           &                      \\
\nii\ $\lambda$6583  & variable &      105  & $+$15~$\pm$10      \\
\enddata

\tablenotetext{a}{The absorption component was measured
from the \ha\ line for the 4 December 2000 observation where bandpass
effects are minimal, the width of the line is not reported because it is
highly dependent on the continuum determination.}
\tablenotetext{b}{Assuming the variable \ha\ component is all due to \ha\ emission.}
\tablenotetext{c}{Assuming the variable \ha\ component is a blend of \ha\ and \ion{He}{2} emission.}

\end{deluxetable}

\end{document}